\documentclass{article}
\usepackage{spconf,amsmath,graphicx}
\usepackage{booktabs,hyperref,subcaption}
\usepackage{dsfont}
\title{Prediction of Satisfied User Ratio for Compressed Video}
\address{$^{\star}$ University of Southern California, Los Angeles, California, USA\\
	 $^{\dagger}$ Netflix, Los Gatos, California, USA\\
	 $^{\ddagger}$ Northwestern Polytechnical University, Xi'an, China.}
\begin{document}
\ninept
\makeatletter
\def\@name{Haiqiang Wang$^{\star}$, Ioannis Katsavounidis$^{\dagger}$,
Qin Huang$^{\star}$, Xin Zhou$^{\ddagger}$, and C.-C. Jay Kuo$^{\star}$}
\makeatother
\maketitle

\begin{abstract}
A large-scale video quality dataset called the VideoSet has been
constructed recently to measure human subjective experience of H.264
coded video in terms of the just-noticeable-difference (JND).  It
measures the first three JND points of 5-second video of resolution
1080p, 720p, 540p and 360p. Based on the VideoSet, we propose a method
to predict the satisfied-user-ratio (SUR) curves using a machine
learning framework. First, we partition a video clip into local
spatial-temporal segments and evaluate the quality of each segment using
the VMAF quality index. Then, we aggregate these local VMAF measures to
derive a global one. Finally, the masking effect is incorporated and the
support vector regression (SVR) is used to predict the SUR curves, from
which the JND points can be derived. Experimental results are given to
demonstrate the performance of the proposed SUR prediction method.
\end{abstract}


\section{Introduction}\label{sec:introduction}

A large amount of bandwidth of fixed and mobile networks is consumed by
real-time video streaming. It is desired to lower the bandwidth
requirement by taking human visual perception into account. Although the
peak signal-to-noise ratio (PSNR) has been used as an objective measure
in video coding standards for years, it is generally agreed that it is a
poor visual quality metric that does not correlate with human visual
experience well \cite{Lin-Kuo-2011}.

There has been a large amount of efforts in developing new visual
quality indices to address this problem, including SSIM \cite{ssim},
FSIM \cite{fsim}, DLM \cite{dlm}, etc. Humans are asked to evaluate the
quality of visual contents by a set of discrete or continuous values called opinion score; typical opinion scores in the range 1-5, with 5
being the best and 1 the worst quality. These indices offer, by definition, users' subjective test results and thus correlate better
than PSNR with their mean (called mean opinion score, or MOS).
However, there is one shortcoming with these indices.  That is, the
difference of selected contents for ranking is sufficiently large for a
great majority of subjects. Since the difference is higher than the
just-noticeable-difference (JND) threshold for most people, disparities
between visual content pairs are easier to tell.

Humans cannot perceive small pixel variation in coded image/video until
the difference reaches a certain level. There is a recent trend to
measure the JND threshold directly for each individual subject.  The
idea was first proposed in \cite{lin2015experimental}.  An assessor is
asked to compare a pair of coded image/video contents and determine
whether they are the same or not in the subjective test, and a bisection
search is adopted to reduce the number of comparisons.  Two small-scale
JND-based image/video quality datasets were built by the Media
Communications Lab at the University of Southern California. They are
the MCL-JCI dataset \cite{jin2016jndhvei} and the MCL-JCV dataset
\cite{mcl_jcv}. They target at the JND measurement of JPEG coded images and
H.264/AVC coded video, respectively.

The number of JPEG coded images reported in \cite{jin2016jndhvei} is 50
while the number of subjects is 30. The distribution of multiple JND
points were modeled by a Gaussian Mixture Model (GMM) in
\cite{hu2015gmm}, where the number of mixtures was determined by the
Bayesian Information Criterion (BIC). The MCL-JCV dataset in
\cite{mcl_jcv} consists of 30 video clips of wide content variety and
each of them were evaluated by 50 subjects. Differences between
consecutive JND points were analyzed with outlier removal. It was also
shown in \cite{mcl_jcv} that the distribution of the first JND samples
of multiple subjects can be well approximated by the normal
distribution. The JND measure was further applied to the HEVC coded
clips and, more importantly, a JND prediction method was proposed in
\cite{huang2017measure}. The masking effect was considered, related
features were derived from source video, and a spatial-temporal
sensitive map (STSM) was defined to capture the unique characteristics
of the source content. The JND prediction problem was treated as a
regression problem.

More recently, a large-scale JND-based video quality dataset, called the
VideoSet, was built and reported in \cite{wang2017videoset}. The
VideoSet consists of 220 5-second sequences, each at four resolutions (i.e.,
$1920 \times 1080$, $1280 \times 720$, $960\times 540$ and $640 \times
360$).  Each of these 880 video clips was encoded by the x264 encoder
implementation \cite{aimar2005x264} of the H.264/AVC standard with
$QP=1, \cdots, 51$ and the first three JND points were evaluated by 30+
subjects.  The VideoSet dataset is available to the public in the IEEE
DataPort \cite{wang2017videoset}. It includes all source/coded video
clips and measured JND data.

In this work, we focus on the prediction of the satisfied user ratio
(SUR) curves for the VideoSet and derive the JND points from the
predicted curves. This is different from the approach in
\cite{huang2017measure}, which attempted to predict the JND point
directly. Here, we adopt a machine learning framework for the SUR curve
prediction. First, we partition a video clip into local spatial-temporal
segments and evaluate the quality of each segment using the VMAF \cite{vmaf} quality
index. Then, we aggregate these local VMAF measures to derive a global
one. Finally, the masking effect is incorporated and the support vector
regression (SVR) is used to predict the SUR curves, from which the JND
points can be derived. Experimental results are given to demonstrate the
performance of the proposed SUR prediction method.

The rest of this paper is organized as follows. The SUR curve prediction
problem is defined in Sec. \ref{sec:satisfied_user_ratio}. The SUR
prediction method is detailed in Sec. \ref{sec:sur_prediction}.
Experimental results are provided in Sec. \ref{sec:experimental_results}. Finally, concluding remarks and future
research direction are given in Sec. \ref{sec:conclusion}.

\begin{figure*}[!t]
\centering
	\begin{subfigure}[b]{1.0\linewidth}
	\centering{}
	\includegraphics[width=0.19\linewidth]{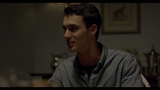}
	\includegraphics[width=0.19\linewidth]{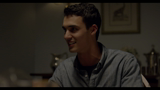}
	\includegraphics[width=0.19\linewidth]{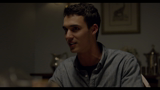}
	\includegraphics[width=0.19\linewidth]{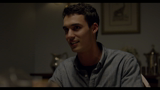}
	\includegraphics[width=0.19\linewidth]{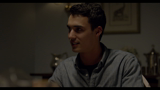}
	\caption{\label{}}
	\end{subfigure}
  \begin{subfigure}[b]{1.0\linewidth}
	\centering
	\includegraphics[width=0.19\linewidth]{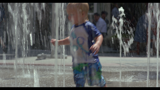}
	\includegraphics[width=0.19\linewidth]{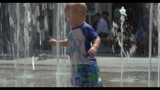}
	\includegraphics[width=0.19\linewidth]{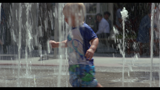}
	\includegraphics[width=0.19\linewidth]{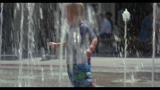}
	\includegraphics[width=0.19\linewidth]{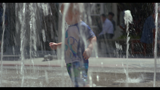}
	\caption{\label{}}
	\end{subfigure}
\caption{Representative frames from source sequences (a) \#37
and (b) \#89.} \label{fig:frames_of_two_sequences}
\end{figure*}

\section{JND and SUR for Coded Video}\label{sec:satisfied_user_ratio}

Given a set of clips $d_{i}$, $i=0, 1, 2, \cdots, 51$, coded from the
same source video $r$, where $i$ is the quantization parameter (QP)
index used in the H.264/AVC. Typically, clip $d_i$ has a higher PSNR value
than clip $d_j$, if $i<j$, and $d_{0}$ is the losslessly coded copy of
$r$.  The first JND location is the transitional index $i$ that lies on
the boundary of perceptually lossless and lossy visual experience for a
subject. The first JND is a random variable rather than a fixed quantity
since it varies with several factors, including the visual content under
evaluation, the test subject and the test environment.
Based on the study in \cite{wang2017videoset}, the JND position can be
approximated by a Gaussian distribution in form of
\begin{equation}\label{eq:jnd_gaussian}
X \sim \mathcal{N} (\bar{x}, s^{2}),
\end{equation}
where $\bar{x}$ and $s$ are the sample mean and sample standard
deviation, respectively.

We say that a viewer is satisfied if the compressed video appears to be perceptually
the same as the reference. Mathematically,
the satisfied user ratio (SUR) of vide clip $d_{i}$ can be expressed as
\begin{equation}\label{eq:sur_sum}
S_{i} = 1-\frac{1}{M}\sum_{m=1}^{M}\mathds{1}_{m}(d_{i}),
\end{equation}
where $M$ is the total number of subjects and $\mathds{1}_{m}(d_{i})=1$
or $0$ if the $m$th subject can or cannot see the difference between
compressed clip $d_{i}$ and its reference, respectively.  The summation
term in right-hand-side of Eq. (\ref{eq:sur_sum}) is the empirical
cumulative distribution function (CDF) of random variable $X$ as given
in Eq. (\ref{eq:jnd_gaussian}). Then, by plugging Eq.
(\ref{eq:jnd_gaussian}) into Eq. (\ref{eq:sur_sum}), we can obtain
a compact formula for the SUR curve as
\begin{equation}
S_{i} = Q(d_{i}|\bar{x}, s^{2}),
\end{equation}
where $Q(\cdot)$ is the Q-function of the normal distribution.

\section{Proposed SUR prediction System}\label{sec:sur_prediction}

\begin{figure*}[!htb]
\centering
\begin{subfigure}[b]{0.48\linewidth}
\centering
\includegraphics[width=1.0\linewidth]{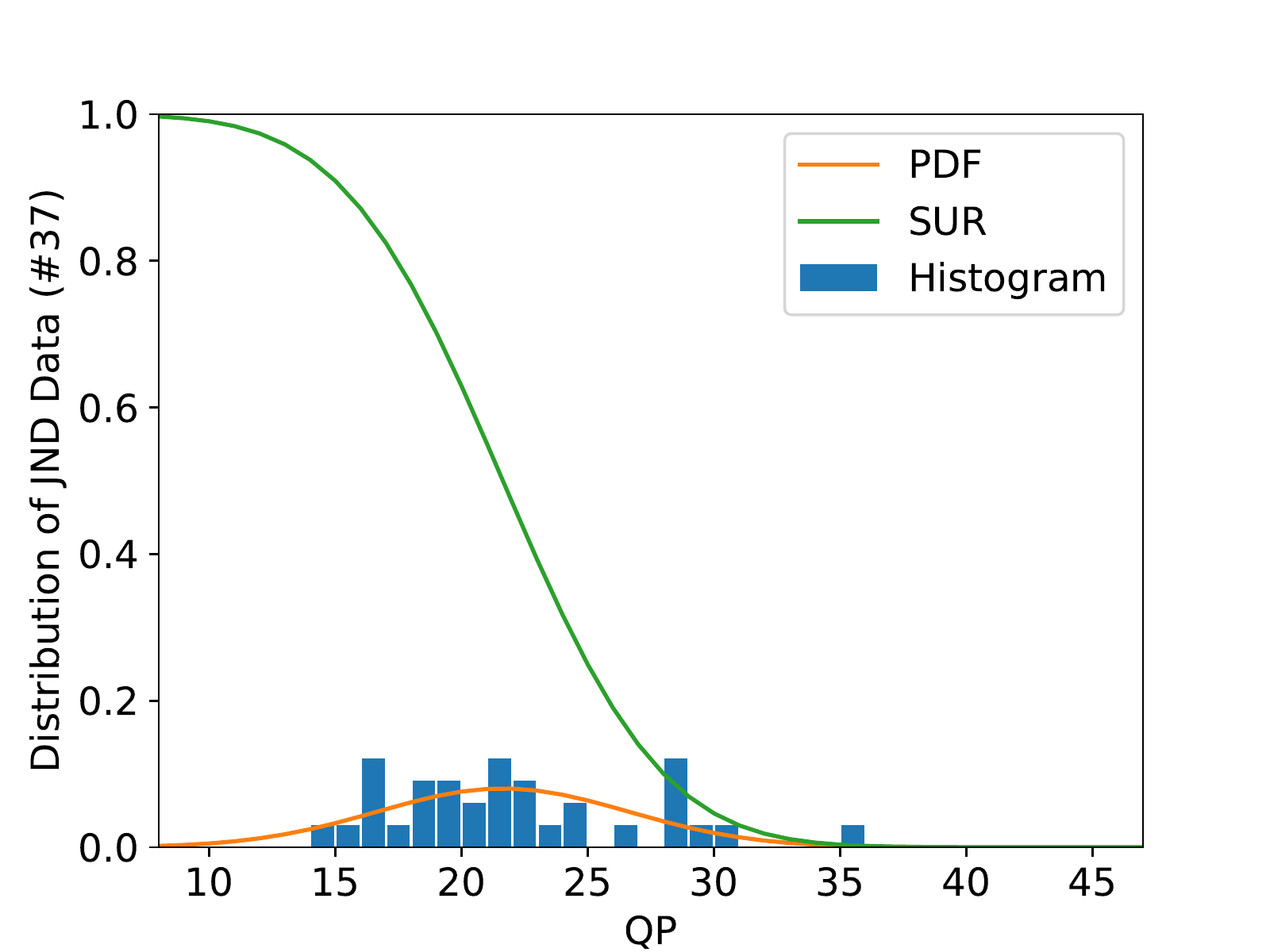}
\subcaption{}
\end{subfigure}
\begin{subfigure}[b]{0.48\linewidth}
\centering
\includegraphics[width=1.0\linewidth]{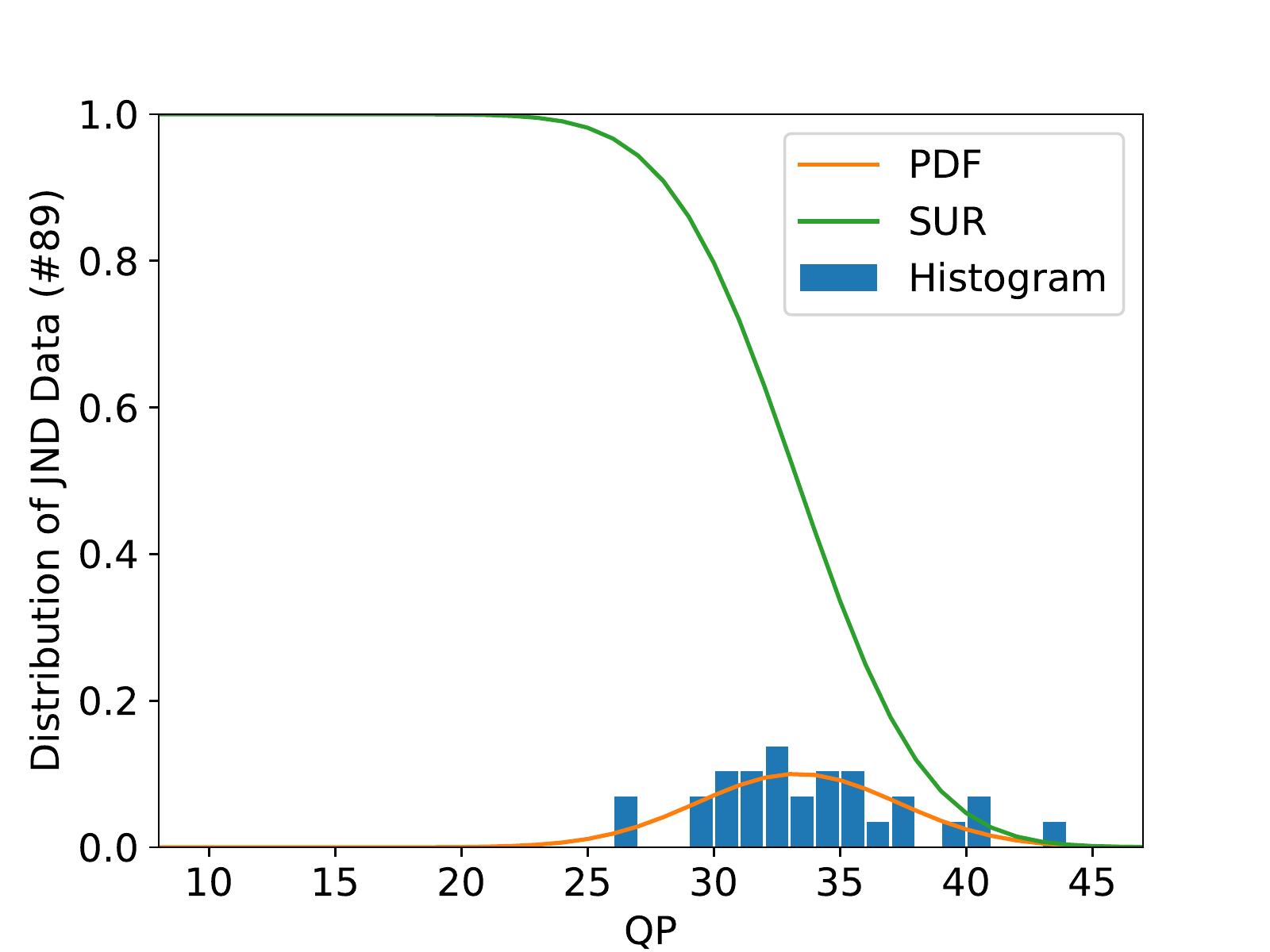}
\subcaption{}
\end{subfigure}
\caption{SUR modeling from JND samples. The JND histogram (in blue), the smoothed PDF curve (in orange)
and the SUR curve (in green) for sequences (a) \#37 and (b) \#89.}\label{fig:data_flow}
\end{figure*}

The SUR curve is primarily determined by two factors: 1) quality
degradation due to compression and 2) the masking effect.  To shed light
on the impact of the masking effect, we use sequences \#37 (DinnerTable)
and \#89 (TodderFountain) as examples. Their representative frames are
shown in Figs. \ref{fig:frames_of_two_sequences} (a) and (b) and their
JND data distributions are given in Figs. \ref{fig:data_flow} (a) and
(b), respectively. Sequence \#37 is a scene captured around a dining
table.  It focuses on a male speaker with still dark background. His face is
the visual salient region that attracts people's attention. The masking
effect is weak and, as a result, the JND point arrives earlier (i.e.  a
smaller $i$ value in $d_i$). On the other hand, sequence \#89 is a scene
about a toddler playing in a fountain.  The masking effect is strong due
to water drops in background and fast object movement.  As a result,
compression artifacts are difficult to perceive and the JND point
arrives later.

\begin{figure}[!t]
\centering
\begin{subfigure}[b]{0.7\linewidth}
\centering{}
\includegraphics[width=1\linewidth]{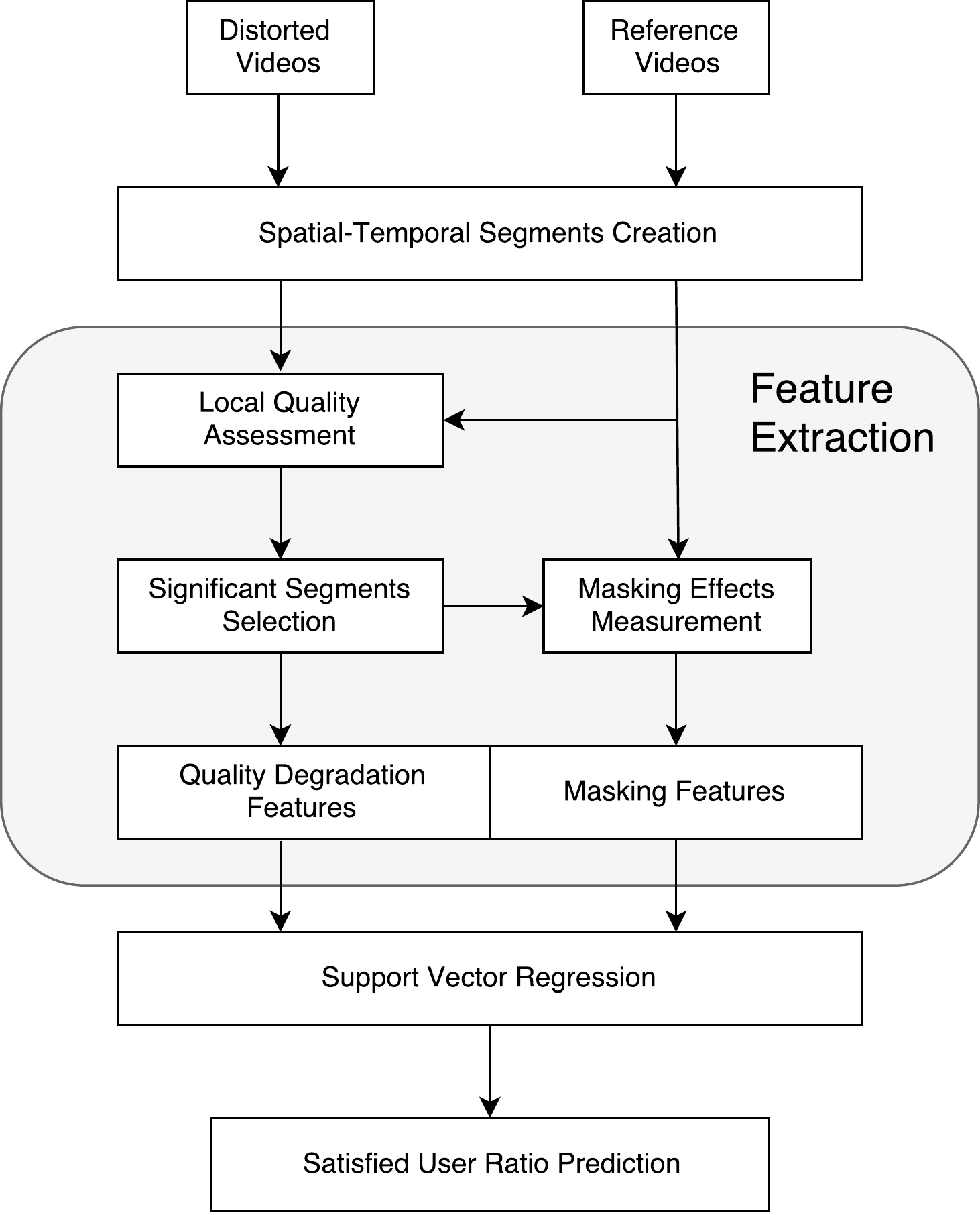}
\end{subfigure}
\caption{The block diagram of the proposed SUR prediction system.} \label{fig:block_diagram}
\end{figure}

The block diagram of the proposed SUR prediction system is given in
Fig.  \ref{fig:block_diagram}.  When a subject evaluates a pair of video
clips, different spatial-temporal segments of the two video clips are
successively assessed. The segment dimensions are spatially and
temporally bounded. The spatial dimension is determined by the area
where the sequence is projected on the fovea.  The temporal dimension is
limited by the fixation duration or the smooth pursuit duration, where
the noticeable difference is more likely to happen than the process of
saccades \cite{hoffman1998visual, ninassi2009considering}.  Thus, the
proposed SUR prediction system first evaluates the quality of local
Spatial-Temporal Segments. Then, similarity indices in these local
segments are aggregated to give a compact global index.  Then, significant
segments are selected based on the slope of quality scores between
neighboring coded clips. After that, we incorporate the masking effect
that reflects the unique characteristics of each video clip. Finally, we
use the support vector regression (SVR) to minimize the $L_{2}$ distance
of the SUR curves, and derive the JND point accordingly. Several major
modules of the system will be detailed below.

\noindent{\bf Step 1. Spatial-Temporal Segment Creation}\label{ssec:spatial_temporal_segments_creation}

The purpose of this module is to divide a video clip into multiple
spatial-temporal segments and evaluate their quality at the eye fixation
level. The dimension of a spatial-temporal segment is $W \times H \times
T$. In case of eye pursuit, the spatial dimension should be large enough
while the temporal dimension should be short enough to ensure that the
moving object is still covered in one segment. In case of eye fixation,
the spatial dimension should not be too large and the temporal dimension
should not be too long to represent quality well at the fixation level.
Based on the study in \cite{wolf2011video,ninassi2009considering}, we
set $W=320$, $H=180$ and $T=0.5s$ here.  The neighboring segments overlap
50\% in the spatial dimension. For example, the original dimension of
720p video is $1280 \times 720 \times 5s$, and there are $7 \times 7
\times 10 = 490$ segments created from each clip.

\noindent{\bf Step 2. Local Quality Assessment}\label{ssec:local_quality_assessment}

We choose the Video Multimethod Assessment Fusion (VMAF) \cite{vmaf} as
the primary quality index to assess quality degradation of compressed
segments. VMAF is an open-source full-reference perceptual video
quality index that aims to capture the perceptual quality of compressed
video.  It first estimates the quality score of a video clip with
multiple high-performance image quality indices on a frame-by-frame
basis. Then, these image quality scores are fused together using the
support vector machine (SVM) at each frame. Results on various video
quality databases show that VMAF outperforms other video quality
indices such as PSNR, SSIM \cite{ssim}, Multiscale Fast-SSIM
\cite{chen2011fast}, and PSNR-HVS \cite{psnr-hvs} in terms of the
Spearman Rank Correlation Coefficient (SRCC), Pearson Correlation
Coefficient (PCC), and the root-mean-square error (RMSE) criteria. VMAF
achieves comparable or outperforms the state-of-the-art video index, the
VQM-VFD index \cite{wolf2011video}, on several publicly available
databases.  For more details about VMAF, we refer interested readers to
\cite{vmaf}.

\noindent{\bf Step 3. Significant Segments Selection}\label{ssec:key_segments_selection}

VMAF is typically applied to all spatial-temporal segments. However,
not all segments contribute equally to the final quality of the entire
clip.  To select significant segments that are more relevant to our
objective, we examine the local quality degradation slope, which is
defined as
\begin{equation}\label{eq:slope}
\delta V(S_{wht}^{d_{i}}) = \frac{V(S_{wht}^{d_{i-k}})-V(S_{wht}^{d_{i}})}{k},
\end{equation}
where $V(S_{wht}^{d_{i}})$ is the VMAF score of segment
$S_{wht}^{d_{i}}$ that is cropped from compressed clip $d_{i}$ with
spatial indices $(w,h)$ and temporal index $t$, respectively.  The slope
in Eq.  (\ref{eq:slope}) evaluates how much the VMAF score of the
current segment $S_{wht}^{d_{i}}$ differs from that in its neighboring
compressed clip $S_{wht}^{d_{i-k}}$, where $k=2$ is the QP difference
between them.  If the slope is small, the local quality does not change
too much and the probability of the associated coding index $i$ to be a
JND point is lower. We order all spatial-temporal segments based on
their slopes and select $p$ percents of them with larger slope values.
We set $p=80\%$ in our experiment. The goal is to filter out less
important segments before we extract a representative feature vector.

\begin{figure*}[!th]
\centering
\begin{subfigure}[b]{0.48\linewidth}
\centering
\includegraphics[width=1.0\linewidth]{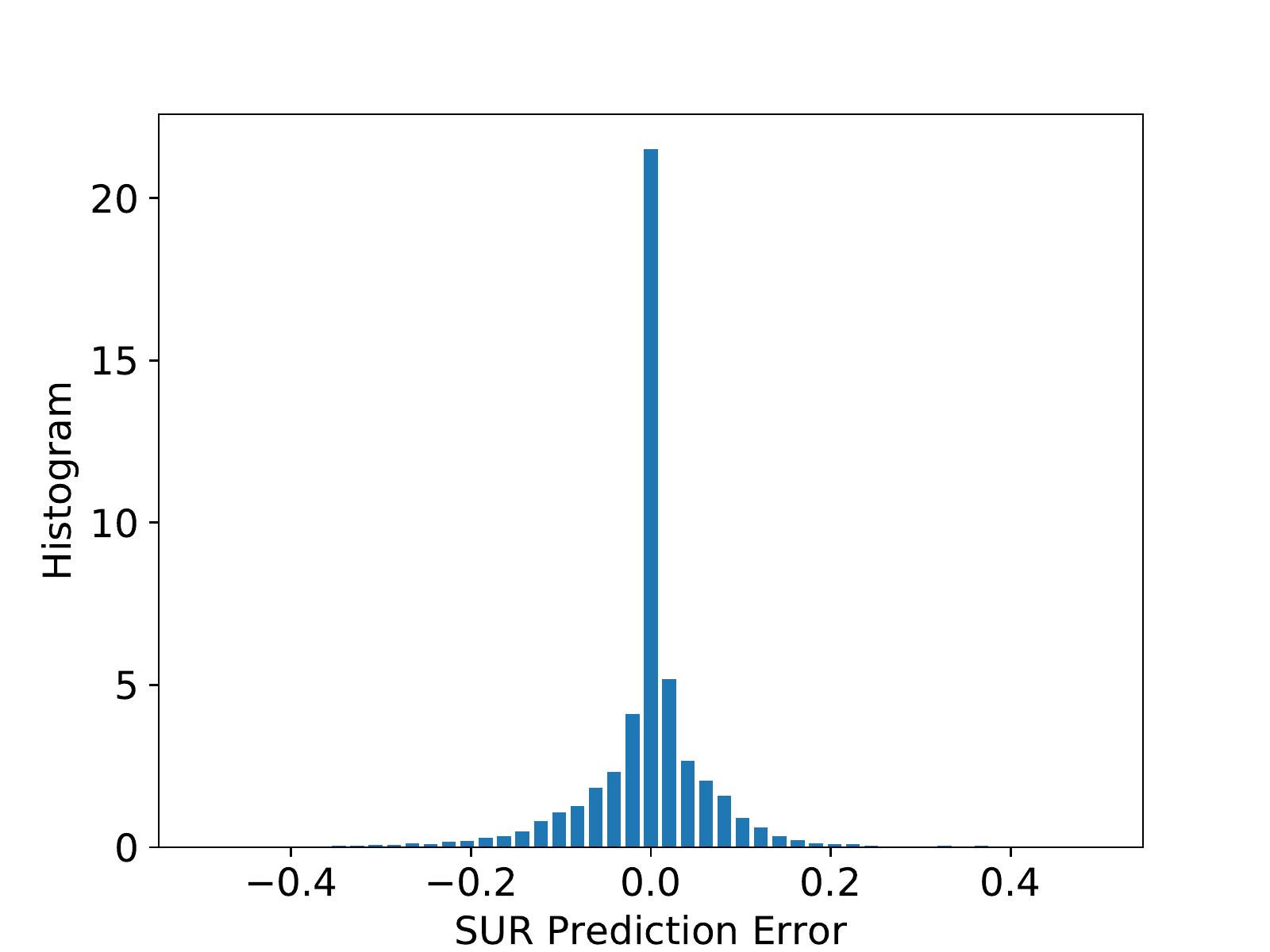}
\subcaption{}
\end{subfigure}
\begin{subfigure}[b]{0.48\linewidth}
\centering
\includegraphics[width=1.0\linewidth]{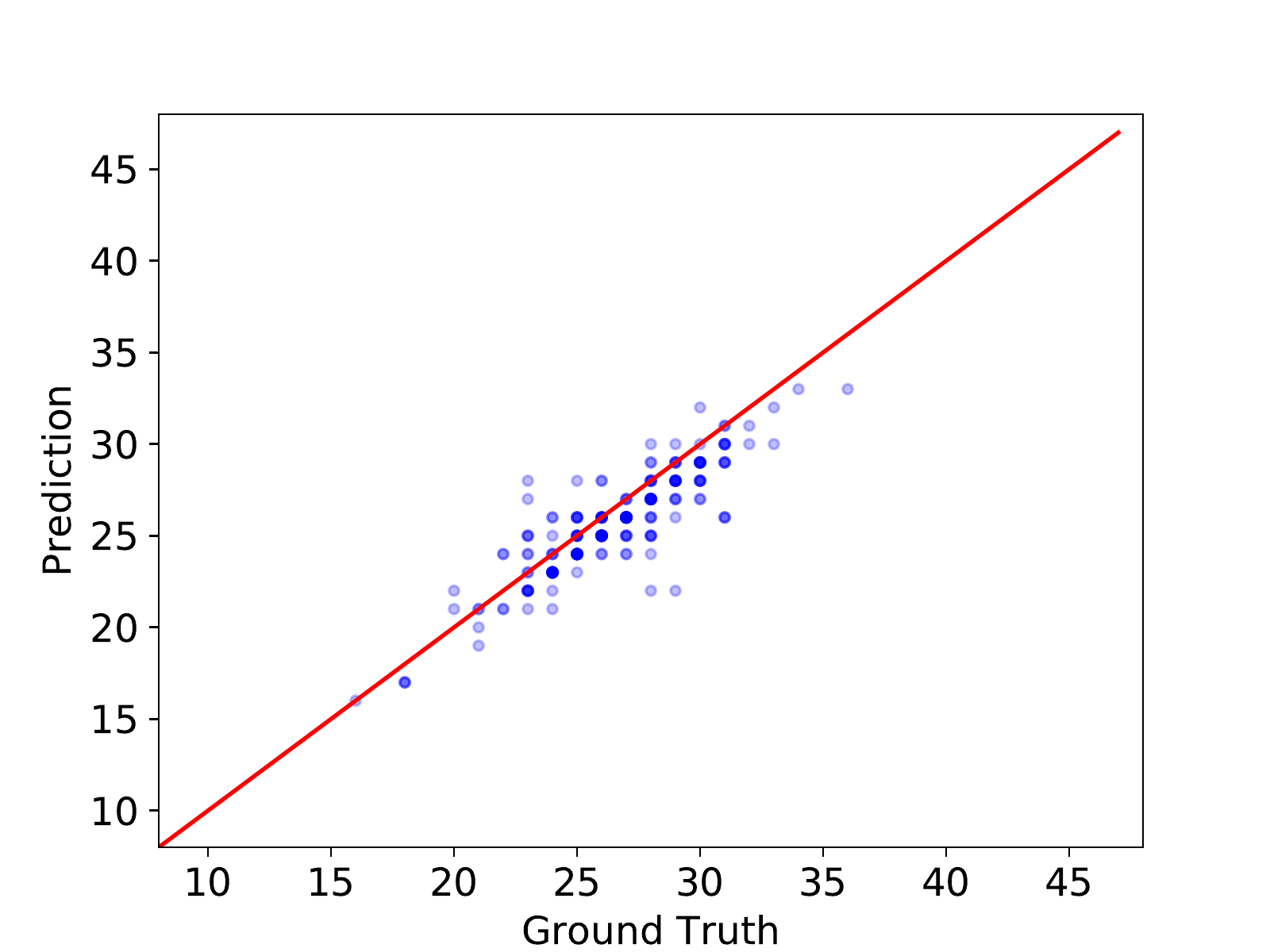}
\subcaption{}
\end{subfigure}
\caption{JND prediction result: (a) the histogram of $\Delta$ SUR
and (b) the predicted VS. the ground truth JND location.}
\label{fig:predicted_error}
\end{figure*}

\noindent{\bf Step 4. Quality Degradation Features}\label{ssec:quality_aware_dynamic_feature}

A cumulative quality degradation curve is computed for every coded clip
based on the change of VMAF scores in significant segments. Its
computation consists of two steps. First, we compute the difference of
VMAF scores between a significant segment from compressed clip $d_{i}$
and its reference $r$ as
\begin{equation}
\Delta V(S_{wht}^{d_{i}}) = V(S_{wht}^{r})-V(S_{wht}^{d_{i}}).
\end{equation}
The values $\Delta V(S_{wht}^{d_{i}})$ collected from all significant
segments can be viewed as samples of a random variable denoted by $\Delta
V(S^{d_{i}})$. Then, based on the distribution of $\Delta V(S^{d_{i}})$,
we can compute the cumulative quality degradation curve as
\begin{equation}\label{eq:curve}
F^{d_i} (n) = Prob[\Delta V(S^{d_{i}}) \le 2n], \mbox{ for } n=1,\cdots,20,
\end{equation}
which captures the cumulative histogram of VMAF score differences for
coded video $d_i$. As shown in Eq. (\ref{eq:curve}), the cumulative
quality degradation curve is represented in form of a 20-D feature
vector.

\noindent{\bf Step 5. Masking Features}\label{ssec:content_aware_static_feature}

As mentioned earlier, quality degradation in a spatial-temporal segment
is more difficult to observe if there exists a masking effect in the
segment. Here, we use the spatial randomness and temporal randomness
proposed in \cite{hu2015tip,hu_objective_2016} to measure the masking
effect. The process is sketched below.  First, high frequency components
of distortions are first removed by applying a low-pass filter, which is
inspired by the Contrast Sensitivity Function (CSF), in the
pre-processing step.  Then, we use the spatial randomness (SR) model
\cite{hu2015tip} and the temporal randomness (TR)
\cite{hu_objective_2016} to compute the spatial and temporal regularity
in a spatial-temporal segment that is generated in Step 1. The spatial
randomness is small in smooth or highly structured regions. Similarly,
the temporal randomness is small if there is little motion between
adjacent frames. When the SR and TR values are higher, the spatial and
temporal masking effects are stronger.  The masking features, $M_{s}$,
are extracted from the reference clip only.
The histograms of the SR and the TR are concatenated to yield the final
masking feature vector:
\begin{equation}\label{eq:mask}
M^{d_0} = [Hist_{10}(SR), Hist_{10}(TR)].
\end{equation}

\noindent{\bf Step 6. Prediction of SUR Curves and JND Points}\label{ssec:sur_curve_regression}

The final feature vector is the concatenation of two feature vectors.
The first one is the quality degradation feature vector of dimension 20
as given in Eq. (\ref{eq:curve}). The second one is the masking feature
vector of dimension 20 as given in Eq.  (\ref{eq:mask}). Thus, the
dimension of the final concatenated feature vector is $40$.  The SUR
prediction problem is treated as a regression problem, and solved by the
Support Vector Regressor (SVR) \cite{smola2004tutorial}.  Specifically,
we adopt the $\epsilon$-SVR with the radial basis function kernel.

\section{Experimental Results}\label{sec:experimental_results}

In this section, we present the prediction results of the proposed SUR
prediction framework. The VideoSet consists of 220 videos in 4
resolutions and three JND points per resolution per video clip. Here, we
focus on the SUR prediction of the first JND and conduct this task for
each video resolution independently. For each resolution, we trained and
tested 220 video clips using the 5-fold validation. That is, we choose
80\% (i.e. 176 video clips) as the training set and the remaining 20\%
(i.e., 44 video clips) as the testing set. We rotated the 20\% testing
set five times so that each video clip was tested once.  Since the JND
location is chosen to be the QP value when the SUR value is equal to
75\% in the VideoSet, we adopt the same rule here so that the JND position
can be easily computed from the predicted SUR curve.

The averaged prediction errors of the SUR curve and the JND position for
video clips in four resolutions were summarized in Table
\ref{tab:perf_comp_sur}. We see that prediction errors increase as the
resolution becomes lower. This is probably due to the use of fixed $W$
and $H$ values in generating spatial-temporal segments as described in
Sec.  \ref{ssec:spatial_temporal_segments_creation}. We will finetune
these parameters to obtain better prediction results in the future.

To see the prediction performance of each individual clip, we use 720p
video as an example. The histogram of the SUR prediction error is given
in Fig. \ref{fig:predicted_error} (a), where the mean absolute error
(MAE) is 0.038 for all test sequences. The predicted JND location versus
the ground-truth JND location is plotted in Fig.
\ref{fig:predicted_error} (b), where each dot denotes one video clip.  As
shown in the figure, most dots are distributed along the 45-degree line,
which indicates that the predicted JND is very close to the ground truth
JND for most sequences.

\begin{table}[!tb]
\centering
\caption{Summary of averaged prediction errors for video clips in four resolutions.}
\label{tab:perf_comp_sur}
\vspace{3mm}
\begin{tabular}{{c}|*{4}{c}} \toprule
              & 1080p & 720p & 540p  & 360p  \\\cline{1-5}
$\Delta$SUR   & 0.039  & 0.038 & 0.037 & 0.042 \\
$\Delta$QP    & 1.218  & 1.273 & 1.345 & 1.605 \\
\bottomrule
\end{tabular}
\end{table}

\section{Conclusion and Future Work}\label{sec:conclusion}

A Satisfied User Ratio (SUR) prediction framework for H.264/AVC coded
video was proposed in this work. It took both the local quality
degradation as well as the masking effect into consideration and extract
a compact feature vector and fed it into the support vector regressor to
obtain the predicted SUR curve. The first JND point can be derived
accordingly.  The system achieves good performance in all resolutions.
We will adopt the same framework to predict locations of the second and
the third JND points in the near future.

\newpage

\bibliographystyle{IEEEbib}
\bibliography{refs}

\end{document}